\PassOptionsToPackage{bookmarksnumbered,unicode,colorlinks=true,linktocpage=true}{hyperref}
\documentclass[sigconf]{acmart}
\usepackage{graphicx}
\usepackage{colortbl}
\usepackage{amsmath}
\usepackage{natbib}
\usepackage{amsfonts}
\usepackage[noend]{algorithmic}
\usepackage[nameinlink,capitalize]{cleveref}
\usepackage{float}
\usepackage{algorithm}
\usepackage{subcaption}
\usepackage{dsfont}
\usepackage{bbm}
\usepackage{multirow}
\usepackage[htt]{hyphenat} %
\usepackage{array}
\usepackage{enumitem}
\setlist[itemize]{leftmargin=4mm}

\usepackage{bibunits}
\defaultbibliography{acm}
\defaultbibliographystyle{ACM-Reference-Format}

\graphicspath{{images//}}
\DeclareGraphicsExtensions{.eps,.png,.pdf,.jpg,.tif,.tiff,.ps}

\newcommand{\pf}[1]{\texttt{#1}}

\usepackage{soul}
\usepackage{xspace}
\definecolor{navy}{rgb}{0.1, 0.1, 0.8}
\definecolor[named]{gray}{rgb}{0.4, 0.4, 0.4}
\definecolor[named]{olive}{rgb}{0.1, 0.5, 0.1}
\definecolor[named]{ruby}{rgb}{0.8, 0.1, 0.3}
\definecolor{darkpastelgreen}{rgb}{0.01, 0.75, 0.24}
\definecolor{celestialblue}{rgb}{0.29, 0.59, 0.82}
\definecolor{coral}{rgb}{1.0, 0.5, 0.31}
\definecolor{Goldenrod}{rgb}{0.8,0.8,0}

\setstcolor{red}
\usepackage[normalem]{ulem}
\newcommand\reduline{\bgroup\markoverwith{\textcolor{red}{\rule[0.5ex]{2pt}{0.4pt}}}\ULon}
\usepackage[colorinlistoftodos,textsize=tiny]{todonotes} %
\newcommand{\eat}[1]{} %

\newcommand{\squishlist}{
\begin{list}{$\bullet$}
 { \setlength{\itemsep}{0pt}
    \setlength{\parsep}{3pt}
    \setlength{\topsep}{3pt}
    \setlength{\partopsep}{0pt}
    \setlength{\leftmargin}{1.5em}
    \setlength{\labelwidth}{1em}
    \setlength{\labelsep}{0.5em} } }

\newcommand{\squishlisttwo}{
\begin{list}{$\bullet$}
 { \setlength{\itemsep}{0pt}
   \setlength{\parsep}{0pt}
   \setlength{\topsep}{0pt}
   \setlength{\partopsep}{0pt}
   \setlength{\leftmargin}{1.5em}
   \setlength{\labelwidth}{1.5em}
   \setlength{\labelsep}{0.5em} } }

\newcommand{\squishend}{
 \end{list}  }

\DeclareMathOperator{\Real}{\mathbb{R}}

\DeclareMathOperator{\His}{\mathcal{H}}

\def\BibTeX{{\rm B\kern-.05em{\sc i\kern-.025em b}\kern-.08emT\kern-.1667em\lower.7ex\hbox{E}\kern-.125emX}}

\newcolumntype{L}[1]{>{\raggedright\let\newline\\\arraybackslash\hspace{0pt}}m{#1}}
\newcolumntype{C}[1]{>{\centering\let\newline\\\arraybackslash\hspace{0pt}}m{#1}}
\newcolumntype{R}[1]{>{\raggedleft\let\newline\\\arraybackslash\hspace{0pt}}m{#1}}

\copyrightyear{2021}
\acmYear{2021}
\setcopyright{acmcopyright}
\acmConference[WSDM '21] {Proceedings of the Fourteenth ACM International Conference on Web Search and Data Mining}{March 8--12, 2021}{Virtual Event, Israel}
\acmBooktitle{Proceedings of the Fourteenth ACM International Conference on Web Search and Data Mining (WSDM '21), March 8--12, 2021, Virtual Event, Israel}
\acmPrice{15.00}
\acmDOI{10.1145/3437963.3441695}
\acmISBN{978-1-4503-8297-7/21/03}

\begin{document}
\fancyhead{}

\newcommand{\titlename}{Birdspotter: A Tool for Analyzing and Labeling Twitter Users}

\title{\titlename}
\author{Rohit Ram}
\affiliation{%
  \institution{University of Technology Sydney}
  \city{Sydney}
  \country{Australia}
}
\email{rohit.ram@uts.edu.au}

\author{Quyu Kong}
\affiliation{%
  \institution{Australian National University \&\\ UTS \& Data61, CSIRO}
  \city{Canberra}
  \country{Australia}}
\email{quyu.kong@anu.edu.au}

\author{Marian-Andrei Rizoiu}
\affiliation{%
  \institution{University of Technology Sydney \& Data61, CSIRO}
  \city{Sydney}
  \country{Australia}
}
\email{marian-andrei.rizoiu@uts.edu.au}

\begin{abstract}
The impact of online social media on societal events and institutions is profound, and with the rapid increases in user uptake, we are just starting to understand its ramifications.
Social scientists and practitioners who model online discourse as a proxy for real-world behavior often curate large social media datasets. A lack of available tooling aimed at non-data science experts frequently leaves this data (and the insights it holds) underutilized.
Here, we propose \pf{birdspotter} -- a tool to analyze and label Twitter users --, and \pf{birdspotter.ml} -- an exploratory visualizer for the computed metrics. 
\pf{birdspotter} provides an end-to-end analysis pipeline, from the processing of pre-collected Twitter data to general-purpose labeling of users and estimating their social influence, within a few lines of code. 
The package features tutorials and detailed documentation.
We also illustrate how to train \pf{birdspotter} into a fully-fledged bot detector that achieves better than state-of-the-art performances without making Twitter API calls, and we showcase its usage in an exploratory analysis of a topical COVID-19 dataset. 
\end{abstract}

\maketitle



\begin{bibunit}

\section{Introduction}

\begin{figure}[tbp]
	\newcommand\myheightA{0.21} %
	\includegraphics[width = 0.49\textwidth]{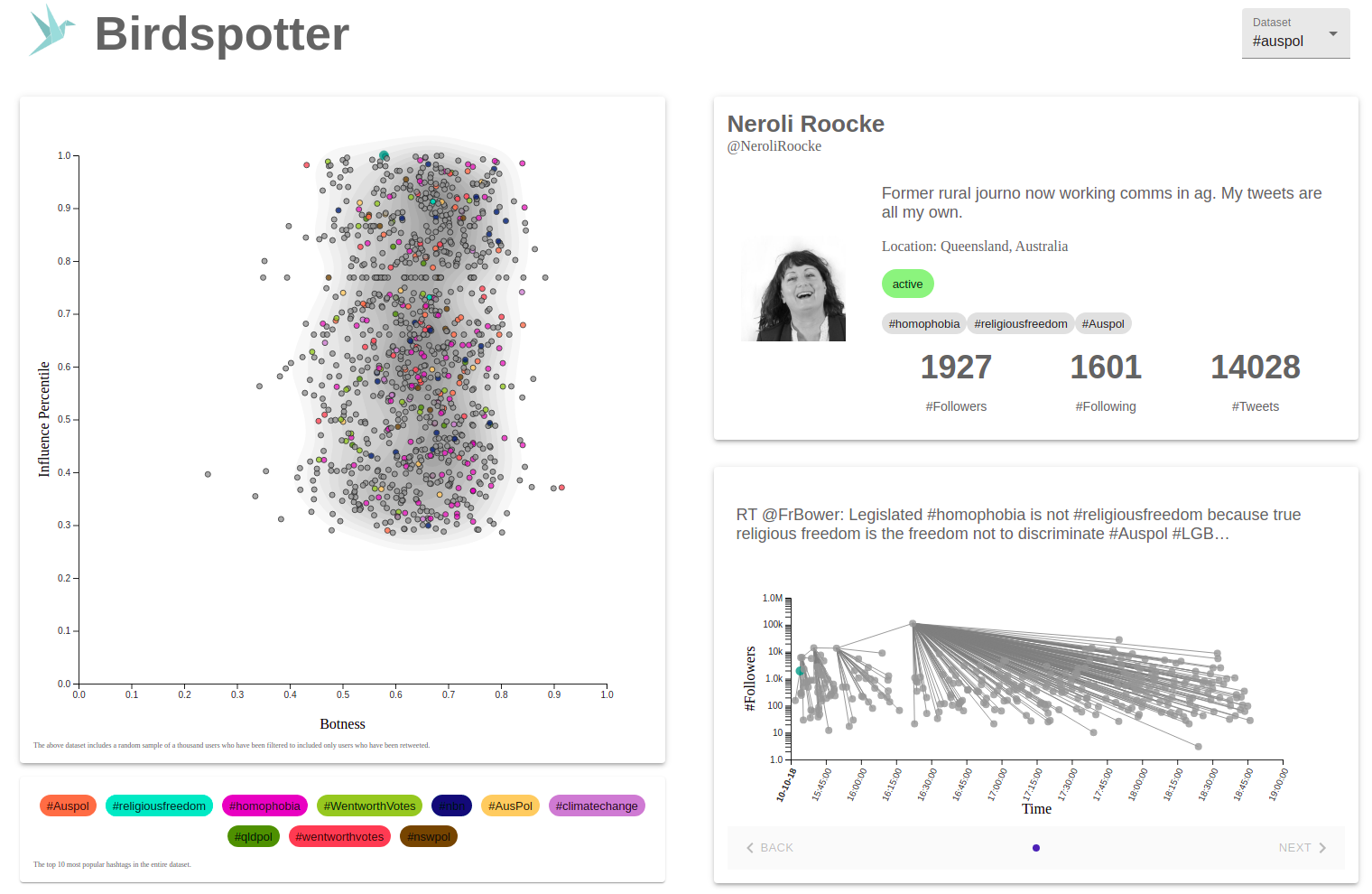}
	\vspace{-7mm}
    \caption{
        The \pf{birdspotter.ml} visualization system: Twitter users are plotted based on their user influence and botness (left panel), and we show a selected user's profile (top-right) and cascade history (bottom-right).
    }
    \label{fig:teaser}
    \vspace{-5mm}
\end{figure}

Barely a decade old, social media in general --- and Twitter in particular --- are becoming increasingly important in shaping societal events.
They serve as novel fora for a wide array of users to express themselves, discuss, promote agendas and attempt to influence the said societal events.
As a result, social and political scientists, journalists, and communication scientists increasingly turn to social media as a proxy to study society.
They carefully curate and label large social media datasets, and here a gap emerges.
There is a limited offer of tools aimed at non-machine learning experts to analyze users in already existing datasets without making additional web API calls that limits the amount of retrieved data.
This paper fills this gap by proposing \pf{birdspotter}, a package aimed at non-computing practitioners with quantitative expertise (basic \pf{R} or \pf{Python}), to analyze, describe, and automatically label users in Twitter datasets.

\addtocounter{footnote}{1}\footnotetext{\label{fn:birdspotter-source}\pf{birdspotter} source code, tutorial, and feature list: \url{https://github.com/behavioral-ds/BirdSpotter}}
\addtocounter{footnote}{1}\footnotetext{\label{fn:birdspotter.ml}\pf{birdspotter.ml} public installation: \url{https://www.birdspotter.ml}}
\addtocounter{footnote}{1}\footnotetext{\label{fn:birdspotter-doc}\pf{birdspotter} documentation: \url{https://birdspotter.readthedocs.io}}
\addtocounter{footnote}{1}\footnotetext{\label{fn:online-tutorial}COVID-19 tutorial: \url{https://github.com/behavioral-ds/user-analysis}}
\addtocounter{footnote}{1}\footnotetext{\label{fn:supp-material}Supplementary Material: \url{https://arxiv.org/pdf/2012.02370.pdf\#page=5}}

This work addresses three specific open questions concerning analyzing Twitter users.
The first question relates to the availability of user analysis tools.
Existing tools are typically designed for Twitter branding and management, i.e. to either analyze a user's or organization's account (Twitter Analytics$^{\ref{fn:twitter-analytics}}$, or Brandwatch Consumer Research$^{\ref{fn:brand-watch}}$), or one given user account (Account Analysis Tool$^{\ref{fn:account-analysis}}$).
The question is \textbf{whether a tool exists to retrospectively analyze and label all the users in Twitter dumps, aimed at non-data science experts with computational expertise?}
We address this question by proposing \pf{birdspotter}$^{\ref{fn:birdspotter-source}}$, an integrated Twitter user analysis tool, that can achieve three types of analysis in only a couple of lines of code.
First, it processes existing Twitter datasets (e.g. \pf{jsonl} data dumps collected through the Streaming API).
Second, it describes users using three types of features (relating to the user, content semantics, and hashtag usage).
Last, it allows training a classifier against a labeled user subset, which turns \pf{birdspotter} into a general-purpose inferential user analysis tool.

\addtocounter{footnote}{1}\footnotetext{\label{fn:twitter-analytics}Twitter Analytics: \url{https://analytics.twitter.com/}}
\addtocounter{footnote}{1}\footnotetext{\label{fn:brand-watch}Brandwatch Consumer Research: \url{https://www.brandwatch.com}}
\addtocounter{footnote}{1}\footnotetext{\label{fn:account-analysis}Account Analysis Tool: \url{https://accountanalysis.app}}

The second open question relates to profiling user botness and influence on previously collected data.
The state-of-the-art bot detector, \textit{botometer}~\citep{sayyadiharikandeh2020detection}, can only be accessed through its web APIs and cannot produce predictions for users that are no longer accessible, such as suspended accounts.
Since bots have a high tendency of being suspended by Twitter, measuring botness a while after collecting data risks missing a large proportion of the bots involved in discussions.
Similarly, existing influence estimation tools require knowledge of the social graph, which often is impossible to capture retrospectively.
The question is: \textbf{can we design a tool that quantifies users' botness and influence on existing curated datasets, without the need of online API calls or supplementary information?}
We address this question two-fold.
First, using four existing Twitter bot datasets, we train \pf{birdspotter} to detect bots without requiring additional API calls.
We show that \pf{birdspotter} achieves a higher performance than the current state-of-the-art \textit{botometer}~\citep{sayyadiharikandeh2020detection};
\pf{birdspotter} ships the bot detector by default, with the package.
Second, we implement a diffusion-based influence estimation~\citep{Rizoiu2017a}, which is as accurate as using the social graph.

The third open question is \textbf{can we visualize and explore both broad and specific views of Twitter users and their activity?}
We address this question by proposing \pf{birdspotter.ml}$^{\ref{fn:birdspotter.ml}}$, a tool that provides both broad views of the user population and detailed inspections of user activity
(see \cref{fig:teaser} for the main interface).

\textbf{The main contributions of this paper are as follows:}
\squishlist
    \item \pf{birdspotter}$^{\ref{fn:birdspotter-source}}$ --- a software package designed for inferential analysis of online users in pre-collected data, and to estimate online user influence based on the reshare cascades.
    \item \pf{birdspotter.ml}$^{\ref{fn:birdspotter.ml}}$ --- an online visualizer designed to perform exploratory analysis of Twitter users.
    \item an offline bot detector, built using four public labeled datasets; we show that it achieves better than state-of-the-art performance and
    we showcase it on an example analysis of users discussing COVID-19$^{\ref{fn:online-tutorial}}$.
\squishend
\noindent{\textbf{Related work.}}
Here, we present the prior work most relevant to \pf{birdspotter}.
For a complete related work discussion, please refer to the online appendix$^{\ref{fn:supp-material}}$.

Tree-based ensemble methods dominate social bot detection (over deep learning) due to the heterogeneity of bots and the relative sparsity of training data. 
The de-facto bot detection tool is \pf{botometer} (formerly \pf{BotOrNot})~\citep{botornot}, which uses more than 1000 user- and recent activity-related features to train a Random Forest classifier.
The main limitations of \pf{botometer} are 1) usage of online APIs which are rate-limited by Twitter, 
2) lack of reproducibility, since deactivated, protected, and suspended users can no longer be retrieved, and 
3) \pf{botometer} scores are likely to vary with user activity and \pf{botometer} versioning.
\pf{Birdspotter} addresses the above by predicting bots on pre-collected Twitter dumps.

User influence is typically measured using static user attributes~\cite{cossu2016review}, analyzing the online social graph~\cite{riquelme2016}, and modeling information diffusion~\cite{zhang2019}.
Closest to our work is \pf{ConTinEst}~\citep{Gomez-Rodriguez2016}, which requires knowledge of the social graph (often prohibitively expensive to obtain) on which it performs random walks (very slow on large social graphs).
\pf{Birdspotter} estimates user influence from resharing dynamics in the absence of knowledge about the social graph.

\section{Preliminaries}
\label{sec:preliminary}
In this section, we briefly outline prerequisites concerning influence estimation using point-process models.
For a thorough construction of the influence estimation, please refer to the online appendix$^{\ref{fn:supp-material}}$.

\noindent{\bf User influence estimation.}
\pf{birdspotter} implements the algorithm in \citep{rizoiu2018debatenight}, estimating online influence as the mean number of retweets generated, directly and indirectly, by a user's (re)tweet. 
\citet{rizoiu2018debatenight} estimate user influence, absent of the retweet branching structure, by assuming that retweets arrive following a Hawkes point process~\citep{Rizoiu2017a}. 
They estimate the probability that the tweet $v_j=(m_j,t_j)$ is a direct retweet of $v_i$ as $p_{ij}=\frac{\phi(m_i,t_j-t_i)}{\sum_{k=1}^{j-1}\phi(m_k,t_j-t_k)}$, where $m_j$ is the associated user's follower count, $t_j$ is the time of the event, and $\phi( m, \Delta t) = \kappa \theta m^\beta e^{-\theta \Delta t}$ is the marked Hawkes exponential kernel of parameters $\kappa$, $\beta$ and $\theta$.
The \emph{pairwise influence} represents the probability that $v_i$ indirectly generates $v_j$, and is computed as 
$r_{ij}=\sum_{k=i}^{j-1}r_{ik}p_{kj}$ when $i < j$, $r_{ii}=1$, and is $0$ otherwise.
Furthermore, a tweet's influence is the sum of its pairwise influences, and a user's influence is its tweets' influences averaged.

\section{Package Overview}
\label{sec:methodology}

In this section, we give an overview of \pf{birdspotter} and \pf{birdspotter.ml}, and describe their usage, functionalities, and design.

\begin{figure*}[!tbp]
    \centering
    \begin{subfigure}{0.34\textwidth}
        \includegraphics[width=\textwidth,page=1]{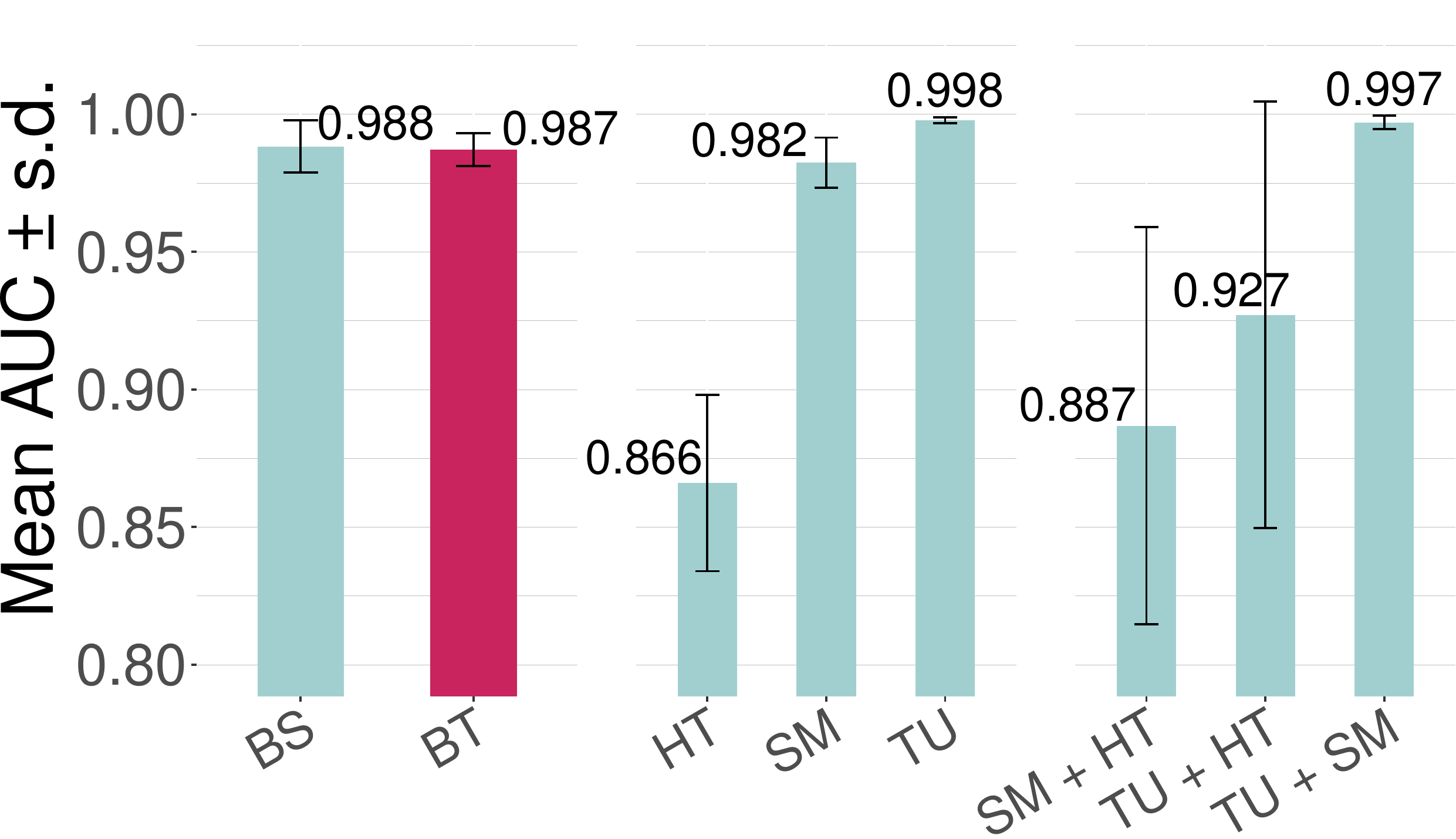}
        \vspace{-5mm}
        \caption{}
        \label{subfig:ablation}
    \end{subfigure}
    \begin{subfigure}{0.34\textwidth}
		\includegraphics[width=\textwidth]{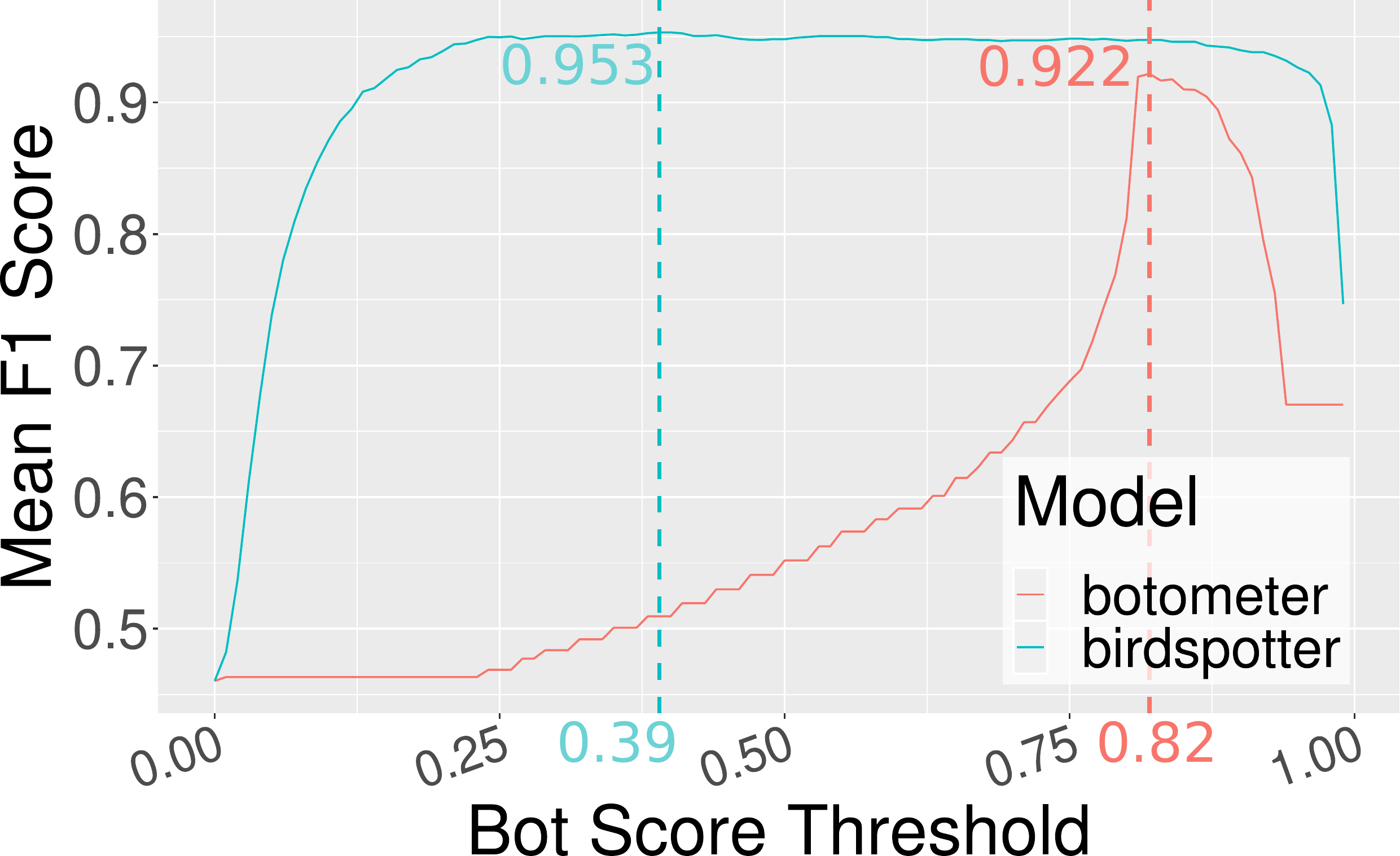}
        \caption{}
        \label{subfig:f1_comparison}
    \end{subfigure}
    \begin{subfigure}{0.3\textwidth}
		\includegraphics[width=\textwidth]{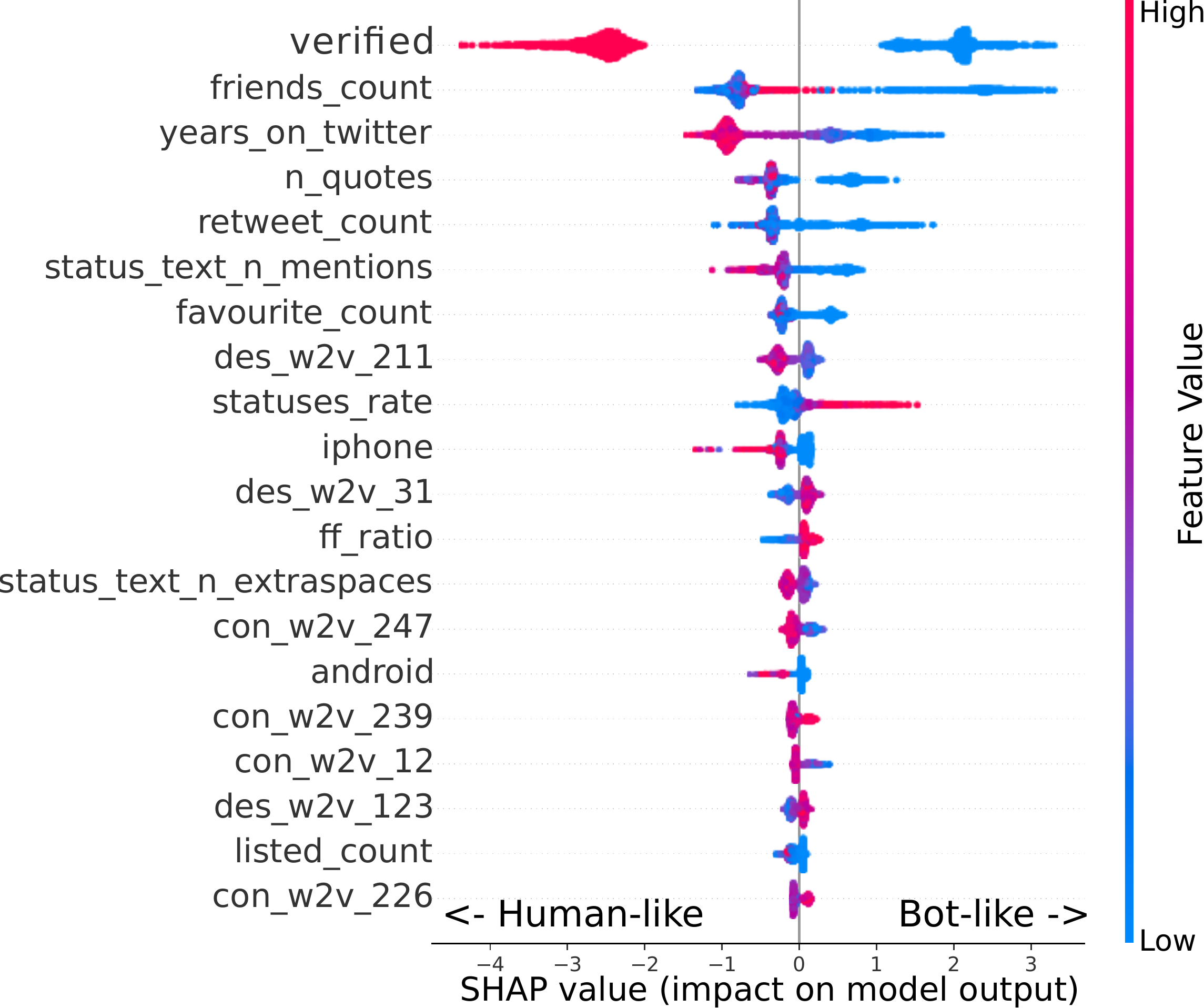}
        \vspace{-5mm}
        \caption{}
        \label{subfig:shap}
    \end{subfigure}
    \vspace{-5mm}
    \caption{
        (a) Mean AUC +/- standard deviation, varying ablated models and botometer. 
        Models/Features are indicated by BS (\pf{birdspotter}), BT (\pf{botometer}), HT (Hashtags), SM (Semantic),and TU (Twitter User).
        (b) Mean $F_1$ score versus bot threshold for \pf{birdspotter} and \pf{botometer}.
        (c) SHAP summary plot where points indicate classifier decisions, y-axis shows features in decreasing importance, x-axis shows SHAP impact value, and color indicates feature value.
        Positive SHAP indicates bots.
    }
    \label{fig:experiments}
    \vspace{-3mm}
\end{figure*}

\vspace*{-0.2cm}
\subsection{\pf{birdspotter}}

\pf{birdspotter} labels users and measures influence
on previously collected tweets in the standard \pf{jsonl} or \pf{json} format. 

\noindent {\bf Measuring influence.} 
\pf{birdspotter} measures user influence as outlined in \cref{sec:preliminary}, using by default a marked Hawkes exponential kernel with parameters $\beta = 1$, $\kappa=\frac{1}{\theta}$ and $\theta = 6.8 \times 10^{-4}$.
These were tuned on a large collection of real cascades~\citep{rizoiu2018debatenight}, and can be customized using the function \pf{getInfluenceScores()}.

\noindent {\bf Usage and functionalities.}
Given a dataset of tweets collected externally (e.g. leveraging the Twitter Filter API), \pf{birdspotter}'s core functionality revolves around two steps.
In the first step, \pf{birdspotter} loads the Twitter dataset, extracts retweet cascades, and compiles the user-level information. 
In the second step, it performs the \textit{influence} analysis and user \textit{labeling}.
The former is achieved by simply invoking the \pf{BirdSpotter} constructor, while the latter is achieved by calling the function \pf{getLabeledUsers()}, which returns a table with the user features detailed above.
For every observed cascade, \pf{birdspotter} also computes the most likely branching structure (see $p_{ij}$ in \cref{sec:preliminary}).
This can be achieved using the function \pf{getCascadesDataFrame()}, which returns the reshare cascades (i.e. original tweet and all its retweets) with the additional column \pf{expected\_parent} indicating a retweet's most likely parent tweet.

For power users, \pf{birdspotter} provides a number of robust configurations --- such as changing the parameters of the Hawkes kernel or using user-defined word embeddings --- documented using its \pf{readthedocs}$^{\ref{fn:birdspotter-doc}}$ documentation. 
A usage tutorial is available on \pf{birdspotter}'s repository$^{\ref{fn:birdspotter-source}}$.
For users who prefer to analyze the results outside \pf{python}, \pf{birdspotter} can dump the user table and the reshare cascades in Comma Separated Values (CSV) files, that can be loaded in outside tools.
All \pf{birdspotter} functionalities can be accessed in \pf{R} via \pf{reticulate} (\url{https://github.com/rstudio/reticulate}).

\noindent{\bf Feature Construction.}
\pf{birdspotter} constructs user features$^{\ref{fn:birdspotter-source}}$ in three categories: Twitter user, semantic, and topic-based features. 
\textbf{Twitter user features} are engineered directly from twitter user attributes and capture heuristics of common bot behavior. 
\textbf{Semantic features} are constructed (by default) from FastText 300d word2vec embeddings~\cite{mikolov2018advances} of users' tweets content and descriptions. 
Content embeddings are averages of tweet embeddings, which are averages of word embeddings.
\textbf{Topic-based features} are the vectors of the 1,000 most frequent hashtags, scored for each user using the term frequency-inverse document frequency scheme.
\pf{birdspotter} is designed to be easily extended with any arbitrary (numerical) features to allow for rapidly evolving bot strategies \citep{yang2019arming}.

\noindent{\bf User labeling.}
\pf{birdspotter} implements a supervised labeler.
It engineers a large selection of features, and it uses a Gradient Boosting Machine model (XGBoost~\cite{xgboost} implementation), with hyperparameters tuned via Random Search and 5-fold cross-validation.

\noindent\textbf{Beyond bot prediction.}
\pf{Birdspotter} ships by default a pre-trained bot classifier (see \cref{sec:experiment}), however \pf{birdspotter} can be customized to a particular application or dataset through labeling and re-training.
The function \pf{getBotAnnotationTemplate()} outputs a CSV that can be annotated by the user, and \pf{trainClassifierModel()} re-trains the classifier with this annotated data.
An option controls whether the model is further tuned starting from the existing model (useful for adapting bot detection to a particular dataset) or retrained from scratch.
We exemplify this in \cref{sec:experiment}.

\noindent {\bf Data Structures.} 
\pf{birdspotter}'s main class, called \pf{BirdSpotter}, is used to access methods and attributes.

\pf{birdspotter} makes accessible three \pf{pandas} dataframes through the main object after processing:
\pf{featuresDataframe} (users and their extracted features), 
\pf{cascadeDataframe} (tweets and cascade information), and 
\pf{hashtagDataframe} (TF-IDF of hashtags).

\noindent {\bf Performance.}
\pf{birdspotter} performed the
extraction, processing, and profiling of a dataset of 196,269 tweets and 129,778 users, in just 5.7 ms per tweet, with an Intel Xeon W-2145 CPU. 

\noindent {\bf Installation.} \pf{birdspotter} installs in the canonical \pf{Python} way: \pf{pip install birdspotter}.

\begin{figure*}[!tbp]
    \centering
    \begin{subfigure}{0.48\textwidth}
        \includegraphics[width=\textwidth]{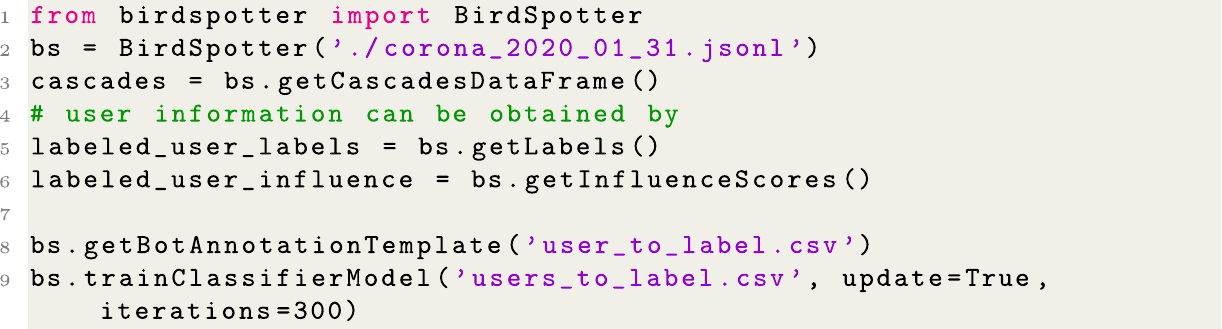}
        \vspace{-5mm}
        \caption{}
        \label{subfig:applications_a}
    \end{subfigure}
    \begin{subfigure}{0.47\textwidth}
		\includegraphics[width=\textwidth]{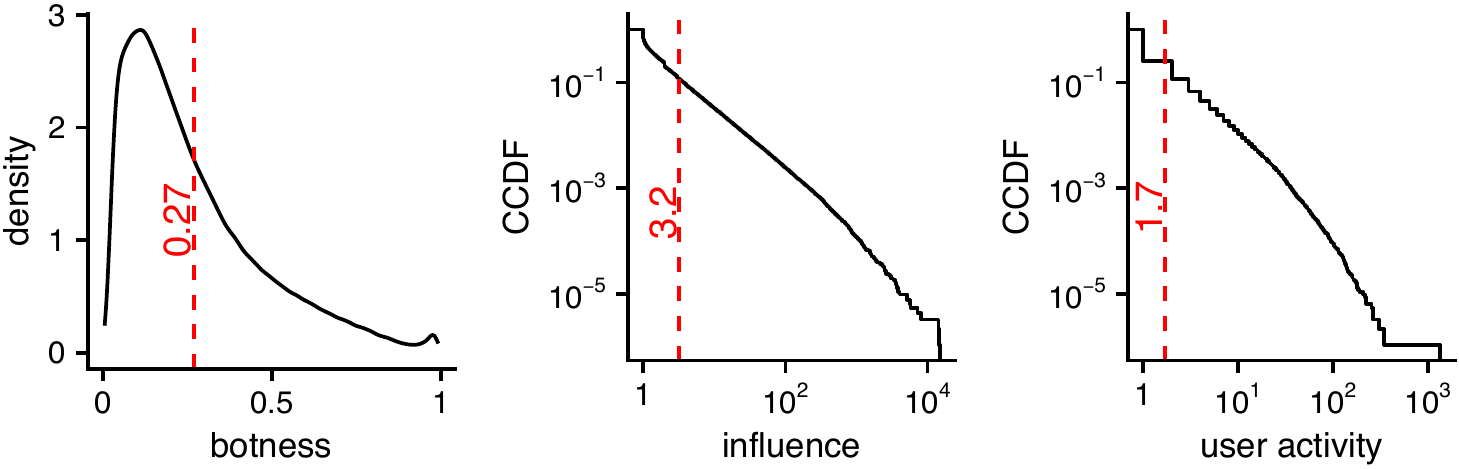}
        \vspace{-5mm}
        \caption{}
        \label{subfig:applications_b}
    \end{subfigure}
    \vspace{-5mm}
    \caption{
        Quantifying user \textit{botness} and \textit{influence} analysis on COVID-19 dataset. 
        (a) Code required to load a Twitter dump, generate cascade and user information, annotate and fine-tune the bot classifier.
        (b) A density plot of user \textit{botness} scores, and complementary cumulative density plots (CCDF) of user \textit{influence} and user \textit{activity}. The  red lines show the mean values.
    }
    \label{fig:dataset-profiling}
    \vspace{-3mm}
\end{figure*}

\vspace*{-.4cm}
\subsection{birdspotter.ml visualiser}
\pf{birdspotter.ml}$^{\ref{fn:birdspotter.ml}}$ is a visualizer built on top of \pf{birdspotter}, and designed to analyze Twitter users engaged in online discussions.
The visualisation provides both broad and specific views of the data, via the three components shown in \cref{fig:teaser}: a scatter plot component, a user information component, and a cascade view component.

\noindent {\bf The Scatter Plot.} 
The left panel contains the scatter-plot showing the influence percentile (on the y-axis) and botness (on the x-axis) of a random sample of the users from the dataset, and the underlying \mbox{2-D} density over the entire data set. 
Users are colored based on the hashtag they use most and, when selected, the user and cascade views are populated.
The plot is pan-able and zoom-able.
The view starts with a random sample of 1,000 users and is dynamically populated 
as practitioners explore cascades.

\noindent {\bf The User View.} The top-right panel provides information about a selected user, including their Twitter image (hyperlink to the user's profile), screen name, location, the hashtags they used, and basic Twitter metrics (such as the number of followers or tweets).

\noindent {\bf The Cascade View.} 
The bottom-right panel shows the cascades the selected user participates in, which are select-able via a carousel.
The component shows the text of the original tweet, the retweets' timing, and the most-likely branching structure inferred using \pf{birdspotter}.
The points on this component are select-able and hover-able in the same way as the scatter plot.
The component also is pan-able and zoom-able.

\section{Building a bot detector} 
\label{sec:experiment}
In this section, we train \pf{birdspotter} as a bot classifier with better performances than the state of the art \pf{botometer}.
We showcase \pf{birdspotter} to profile a topical COVID-19 Twitter dataset.

\noindent{\bf Training data.}
\pf{birdspotter} provides the functionality to retrain and update the current model, which we leverage to build a bot detector.
We train on four public bot datasets, including 
$\{$\pf{botometer‐feedback-2019}, \pf{political-bots-2019}$\}$~\citep{yang2019arming}, and 
$\{$\pf{verified-2019}, \pf{botwiki-2019}$\}$~\citep{yang2020scalable}, sourced from \emph{Bot Repository}\footnote{available from \url{https://botometer.osome.iu.edu/bot-repository/datasets.html}}.

\noindent{\bf Training.}
The \emph{Bot Repository} only provides account-level data, whereas \pf{birdspotter} is designed to utilize tweet \pf{jsonl}. We use the tool \pf{twarc} to acquire the timeline of each available user's first 200 tweets, to construct \pf{jsonl} training data.
We extract and preprocess the data with \pf{BirdSpotter()}, label the resulting dataframe with users' ground truth values, and run \pf{trainClassifierModel()} on this training data to acquire our final model.
We ship this model as the default at \pf{birdspotter}'s installation.

\noindent{\bf Botometer comparison.}
We compare the derived model against \pf{botometer}, by acquiring their bot scores (universal CAP \citep{yang2019arming}) for available users through their API.
\cref{subfig:ablation} shows that \pf{birdspotter} out-performs \pf{botometer} in terms of mean AUC, despite using less information to make predictions -- \pf{botometer} uses more user features extracted from the online API.
\cref{subfig:f1_comparison} shows that \pf{birdspotter} consistently out-performs \pf{botometer} with respect to mean $F_1$ scores, over all bot score thresholds.

\noindent{\bf Ablation study.}
We test the importance of each set of features through various ablations of our main model. 
\cref{subfig:ablation} shows the mean AUC obtained for subsets of features. 
It shows that Twitter user features and semantic features are both informative of bot-like behavior, while hashtag features show more variation. 
The hashtag model performance may be an artifact of training on the mixture of bot datasets (containing hashtags relating to different topics). 
We retain hashtag features in \pf{birdspotter}, for better generalizability when users train and test on their own domain datasets. 
The best performing model uses Twitter user features and semantic features.

\noindent{\bf SHAP analysis.}
We use \pf{shap} \citep{NIPS2017_7062} for explaining the impact of features in our tree ensemble model.
\cref{subfig:shap} shows that the Twitter user features form the majority, and semantic features a minority of the impactful features, in line with the ablation study.

\noindent{\bf COVID-19 Application Dataset.}
We apply \pf{birdspotter} to a COVID-19 dataset~\citep{chen2020tracking}, supplied as tweet IDs which were \textit{re-hydrated} with \pf{twarc} to a \pf{jsonl} format, recovering $68.8\%$ . We limit our analysis to the $\sim 1.5$M unique tweets relating to posts on January the 31st, resulting in $\sim 0.28$M users and $\sim 0.42$M cascades.

\noindent {\bf Dataset profiling.}
\cref{subfig:applications_b} shows the empirical distributions of botness, influence, and activity (i.e., the number of cascades a user participates in). The distribution of botness indicates two maxima; the larger indicating the humans and the smaller indicating the bots . 
Conforming with the literature~\cite{rizoiu2018debatenight}, influence and activity are long-tailed (following a ``rich-get-richer'' paradigm).

\noindent {\bf (Re-)Labeling Users.} 
Exploring \pf{birdspotter.ml} we observe humans --- \pf{@DumplingSays}, \pf{@eddfuentess}, and \pf{@marat\_dospolov} --- with bot scores of $0.873$, $0.83$, and $0.925$ respectively.
Using \pf{getAnnotationTemplate} (see \cref{subfig:applications_a}, line 8) we label each user as \emph{human}, and update the classifier with \pf{trainClassifier} (\cref{subfig:applications_a}, line 10). 
The new bot scores are $0.375$, $0.296$, and $0.559$, respectively.
Practitioners can use \pf{birdspotter} to classify any latent user attribute.

\section{Conclusion}
We presented \pf{birdspotter}, a Twitter user analysis tool aimed at non-data science experts who analyze discourse and user activity on social media. 
It provides an end-to-end analysis of users' online characteristics, and populates a visualizer facilitating both broad views of a user population and individual exploration.
As with many open-source classifiers, we know that \pf{birdspotter} could be leveraged to infer sensitive features.
However, we are currently not aware of any protections that we could implement to prevent this.

Tools like \pf{birdspotter} are integral to the timely, performant, and reproducible analysis of social media users for understanding discourse and society.

\vspace*{-1.5mm}
\subsection*{Acknowledgments}
\small{This research was partially funded by Facebook Research under the Content Policy Research Initiative grants and the Defence Science and Technology Group of the Australian Department of Defence.
} 

\vspace*{-1.5mm}
\putbib
\end{bibunit}

\begin{bibunit}
\clearpage
\appendix

Accompanying the submission \textit{\titlename}.

\section{Additional Related Work}
In this section, we outline other approaches to bot detection and influence measurement in the literature.

\noindent {\bf Detecting Twitter bots.}
There have been a myriad of approaches to detect bots on Twitter. 
There are three motifs within the literature.
The first motif are supervised methods used to determine if an individual user is a bot, usually employing feature construction.
Such approaches include NLP approaches \citep{knauth2019, clark2016}, deep-learning approaches \citep{kudugunta2018}, feature-engineering \citep{chu2013, yang2014, botornot} and other methods \citep{mazza2019, ferrara2016}.
The second motif are unsupervised methods used to discover coordinated online behavior/real-time online campaigns; and 
the third motif are adversarial methods which achieve better bot detection by generating better bots. 

\pf{birdspotter} falls in the first category. It uses a supervised approach to retrospectively analyze datasets. 
It satisfies a different use case than coordinated online behavior tools like \pf{BotSlayer}~\citep{botslayer}. 
Adversarial approaches are fairly novel, however it is unclear whether they might simply improve bot technology, as they provide recipes to build better bots.

The de-facto bot detection tool in the social science community is \pf{Botometer} (formerly \pf{BotOrNot})~\citep{botornot}, which uses more than 1000 user- and recent activity-related features to train a Random Forest classifier.
\pf{Botometer} is currently at version 4, at the time of writing, and serves half a million queries a day~\citep{sayyadiharikandeh2020detection}. 

The main limitation of \pf{botometer} for practitioners is its dependence on an online API. 
It cannot be used to profile the users in offline Twitter datasets which have been collected in the past (like used in many works~\cite{wojcik2018, bessi2016, ferrara2020}). 
Furthermore, the API is rate-limited by Twitter, and requires registration through both Twitter and \emph{RapidAPI} service. For scientific purposes, \pf{botometer} makes local reproducibility difficult to achieve, since deactivated, protected, and suspended users can no longer be retrieved, and \pf{botometer} scores are likely to vary with user activity and \pf{botometer} versioning.

\pf{Birdspotter} addresses the above-stated shortcomings by producing bot predictions on already collected Twitter dumps, and  exposing a simple interface to allows researchers to annotate their own Twitter user collection.

\textbf{Tools for quantifying online influence.}
There are many features used to score the influence, reputation or popularity of online users. 
We delineate these into three areas: those using static user attributes (including lexical features and information on a user's profile)~\cite{cossu2016review}, those that analyze the online social graph (e.g. degree, PageRank, HITs, etc.)~\cite{riquelme2016,cha2010}, and those modeling information diffusion~\cite{zhang2019}.
However, few of these have translated into accessible tools for the non-experts in the field.
For instance, \citet{cossu2016review} provide a set of scripts to perform their influence measurement method.
Other tools, like \pf{ConTinEst}~\citep{Du2013,Gomez-Rodriguez2016}, require knowledge of the social graph (which is often prohibitively expensive to obtain) on which it performs random walks (which are very slow on large social graphs).
\pf{Birdspotter} estimates user influence from reshare dynamics, in the absence of knowledge about the social graph, and provides an end-to-end tool to analyze Twitter users.

\section{Influence measure}
We review the theoretical prerequisites concerning modeling reshare cascades using point processes, and estimating reshare influence.

\noindent{\bf Reshare cascades.}
\pf{birdspotter} analyzes the spread of online information in the form of online \textit{reshare cascades}. 
A reshare cascade consists of an initial user post and some reshare events of the post by other users. 
On Twitter, for example, this can happen when users use the retweet functionality.
We denote a cascade observed up to time $T$ as $\His(T) = \{t_0, t_1, \dots\}$ where $t_i \in \His(T)$ are the event times relative to the first event ($t_0 = 0$). 
We denote cascades with additional information about events --- dubbed here as \emph{event marks} --- as marked cascades.
We use the notation $\His_m(T) = \{(t_0, m_0), (t_1, m_1), \dots\}$, where each event is a tuple of the event time and the event mark. 
For example, for retweet cascades, the numbers of followers of a Twitter user are commonly adopted as event marks~\citep{Zhao2015SEISMIC:Popularity,Mishra2016FeaturePrediction,Mishra2018ModelingPopularity}.

\noindent {\bf The Hawkes processes.} 
\pf{birdspotter} models reshare cascades using Hawkes processes~\citep{hawkes1971spectra} --- a type of point processes with the self-exciting property, i.e.,
the occurrence of past events increases the likelihood of future events. 
The occurrence of events in a Hawkes process is controlled by the event intensity function:
\begin{equation}
    \lambda(t \mid \His(T)) = \mu(t) + \sum_{t_i < t} \phi(t - t_i)
\end{equation}
where $\mu(t)$ is the background intensity function and $\phi: \Real^+ \rightarrow \Real^+$ is a kernel function capturing the decaying influence from a historical event. 
We note that, for reshare cascades, all events are considered to be offspring of the initial event, i.e. there is no background event rate $\mu(t)=0$. Two widely adopted parametric forms for the kernel function $\phi$ include the exponential function $\phi_{EXP}(t) = \kappa \theta e^{-\theta t}$ and the power-law function $\phi_{PL}(t) = \kappa (t + c)^{-(1+\theta)}$.

\noindent {\bf Marked Models.} 
\pf{birdspotter} implements marked versions of the point processes, where the mark is the number of followers that the user emitting the tweet has.
This is because the mark of each event governs the number of future events, e.g., a tweet from a largely followed user is likely to attract more retweets.
The marked versions of Hawkes processes~\citep{Mishra2016FeaturePrediction} are then derived by rescaling the kernel functions with the marks, i.e., $\phi(m, t) = m^\beta \phi(t)$;
$\beta$ controls the warping effect of the mark.

\noindent{\bf User influence estimation.}
\pf{birdspotter} adopts the following definition for user influence, widely used in literature~\citep{Du2013,Zarezade2017,rizoiu2018debatenight}:
\theoremstyle{definition}
\begin{definition}{Online user influence $\varphi(u)$}
	is defined as the mean number of reshares generated directly and indirectly by a message posted by $u$, irrespective if it is an original message or a reshare.
\end{definition}

\noindent Estimating influence from retweet cascades has the additional difficulty of not observing the branching structure of the diffusion --- i.e., the Twitter API attributes all retweets to the original tweet.
\pf{birdspotter} estimates Twitter user influence using only the observed retweet cascade $\His_m(T) = \{v_0 = (t_0, m_0), v_1 = (t_1, m_1), \dots\} $, where marks correspond to users' number of followers. 

\citet{rizoiu2018debatenight} propose a method to estimate user influence in the absence of the branching structure by assuming that retweets arrive following a Hawkes point process~\citep{Rizoiu2017a}.
We can quantify the probability that an event $v_j$ is generated by a previous event $v_i$ as the ratio of the event intensity generated by $v_i$ and the total intensity at time $t_j$. 
Formally, the probability $v_j$ retweets $v_i$ is 
\begin{equation}
p_{ij}=\frac{\phi(t_j-t_i)}{\sum_{k=1}^{j-1}\phi(t_j-t_k)}
     \label{eq:pij}
\end{equation}

\citet{rizoiu2018debatenight} also introduce the pairwise influence score $m_{ij}$, intuitively defined as the amount of \emph{influence} that $v_i$ exerts over $v_j$ either directly (when $v_j$ is a direct retweet of $v_i$) or indirectly (when $v_j$ is a retweet of a descendant of $v_i$):
\begin{equation}
m_{ij}=
\begin{cases}
\sum_{k=i}^{j-1}m_{ik}p_{kj} &, i \leq k < j\\
1 &,i=j\\
0 &,i>j
\end{cases} \enspace,
\label{eq:mij}
\end{equation}

Finally, the influence of $v_i$ is $\varphi(v_i)=\sum_{k=i}^nm_{ik}$, and 
the influence of a user $u$ is the average of the influences of all of their tweets:
\begin{equation}
    \varphi(u)=\frac{\sum_{v\in\mathcal{T}(u)}\varphi(v)}{|\mathcal{T}(u)|}
\end{equation}
where $\mathcal{T}(u)$ is the set of all the tweets emitted by user $u$.
\putbib
\end{bibunit}
\end{document}